# Spectroscopy of deprotonated anions of cold chlorophyll pigments, an approach to their photosynthetic properties.

A. Muheddine[1], B. Robert[2], N. Shafizadeh[1], B. Soep[1]

*[1] Université Paris-Saclay, CNRS, Institut des Sciences Moléculaires d'Orsay (ISMO), Batiment 520 Rue André Rivière 91405, Orsay, France*

.

*[2] Université Paris-Saclay, CEA, CNRS, Institute for Integrative Biology of the Cell (I2BC), 91198, Gif-sur-Yvette, France.*

## Abstract

We introduce here a new approach to the study of isolated chlorophyll pigments with the electronic spectroscopy of their deprotonated anions in a cold environment at 9K. These anions are isoelectronic to the corresponding neutral molecules, therefore their electronic properties are comparable. This comparison is pursued through this article. The electronic spectrum of the Q bands of deprotonated pheophorbide and methyl pheophorbide are recorded by resonant two photon electron detachment with vibrational resolution and show well-resolved bands extending through the visible spectrum in a very similar fashion to their neutral counterparts in solution. The results are compared with data of neutral pheophytin obtained by fluorescence line narrowing spectroscopy at 4K and resonance Raman spectroscopy and found very similar. The theoretical analysis is achieved by density functional (ground state DFT ωB97XD, hybrid exchange-correlation functionals and 6-311++G(d,p) basis sets) and time dependent density functional (TDDFT) calculations. The comparison of the active orbitals in the Q transitions shows a great similarity between neutral and anionic forms together with some orbital and related level inversions. The vibronic spectra are simulated by Franck-Condon calculations to assign the dominant transitions of the electronic spectrum. Globally, this is a new perspective in the electronic spectroscopy of chlorophyll pigments that has been opened.

## Introduction

The heart of photosynthesis is a charge separation process in a reaction centre activated by light energy collected by antennas. Chlorophyll a pigments are at the centre of this process as they perform most of the light collection and excitation energy transfer in the antenna proteins, as well as charge separation and the first electron transfers in photosystems I and II, ultimately resulting into production of ATP and NADPH, the two cofactors necessary for the reduction of carbon dioxide into sugars. In the reaction centres the charge separation and the subsequent electron transfers occur with a surprisingly high quantum yield, close to unity. In photosystem II charge separation occurs in a special pair of chlorophylls and is rapidly transferred to a pheophytin (a demetalated chlorophyll), through a neighbouring chlorophyll a. There has been in the recent years a series of thorough measurements on reaction centers by 2D spectroscopy and its variants to unravel the reaction paths of primary charge separation in photosystem II [1-3] that underscored the absorption by exciton states of the chlorophyll special pair, a chlorophyll and a pheophytin. While a reaction scheme emerges with well identified steps[3], modelling is hampered by the lack of precise experimental data on the spectroscopy of chlorophylls. The precise mechanism of charge separation in impulse driven experiments is still under examination. A possible mechanism involves a low frequency vibrational mode would drive an exciton pair to a charge separated state through a resonance with a high efficiency,[1,4] as required in natural systems.

Through many years, it was understood that spectroscopy is essential for deciphering this mechanism[5-12] inspecting vibrational electronic structure or vibronic couplings. While resonance Raman has been a tool of choice to detail the electronic ground state vibrational structure of chlorophyll systems within reaction centres and antennas, the information on the detailed vibronic structure of the first excited electronic states of chlorophyll pigments is scarce and has been up to now addressed experimentally by only two groups in solid solutions at 4K by site selective spectroscopy [8,13] and by spectral hole burning [11,12,14]. These experiments yielded a body of information on the vibronic properties of the ground- and first electronically excited states of chlorophylls in biologically active media. Several experiments have tracked the spectroscopy of chlorophyll in the gas phase in neutral chlorophyll[15] and chlorophyll tagged with ions positive or negative[16-18].

However, the latter tagging method fails to capture the low frequency vibrations, while matrix measurements (site selective or hole burning) probe pigments 5 or 6-coordinated with a solvent.

We endeavour here to use a new methodology in the gas phase that will capture the whole visible electronically excited spectrum of chlorophyll pigments devoid of interaction with solvents, given the fact that the solvent interactions are important for chlorophylls[19,20]. To this end, we study here deprotonated anions of chlorophyll pigments which bear a negative charge away from the chlorophyll core orbitals as depicted for example by Tsujimura et al[21], rationalising observations of deprotonated anions of chlorophyll $C_1$ and $C_2$ in diatom algae [22,23]. In effect, the spectra of these chlorophylls are strikingly similar in their protonated and deprotonated forms despite the fact that the acrylate group at position 7 is conjugated to the porphyrin cycle ( see Figure *1* for the generic pigment depiction). This shows that studying deprotonated chlorophyll pigments may constitute a way of addressing their neutral state properties in the gas phase.  Although the involvement of deprotonated chlorophyll c in the antennas of diatoms *in vivo* is not formally assessed, it suggests the potential importance of negative ions in photosynthetic mechanisms.

The essential pigments in the primary reaction centres are chlorophylls and pheophytin a, a demetalleted chlorophyll a. In this paper, we concentrate our studies on the latter molecule as deprotonated anion since the vibronic coupling between Qx and Qy excited states is weak and the states well separated. This will allow applying the findings to the more overlapped chlorophyll transitions[5]. We report here the first observation of cooled, vibrationally resolved electronic spectra of the first electronic transitions of pheophorbide and methyl  pheophorbide in their anionic deprotonated states. These molecules are closely related to pheophytin a, with only modifications on the non-conjugated phytyl chain of pheophytin a (Figures 1 a,b).

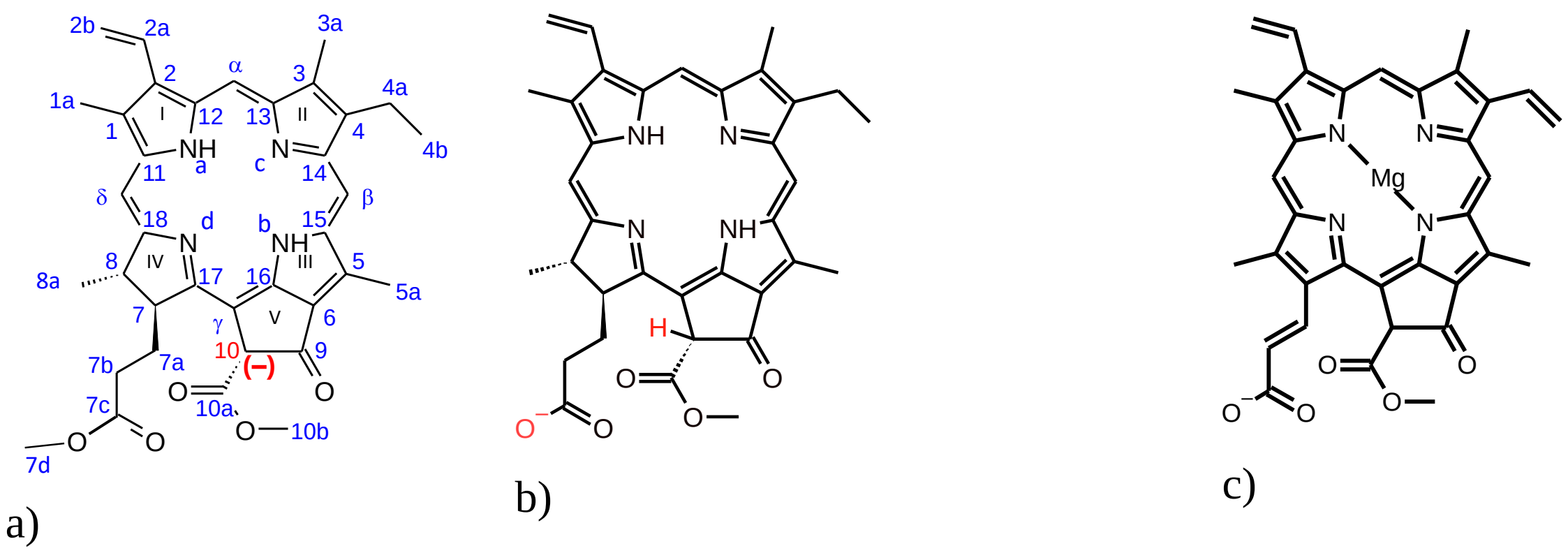

Figure 1 : structures of the deprotonated methyl  pheophorbide a), deprotonated  pheophorbide b), deprotonated chlorophyll C2 c).

The latter anions are *isoelectronic* to the respective neutrals pheophorbide, methyl pheophorbide and Chlorophyl C2 through the removal of a proton. However the extra charge is located in different positions in these molecules as documented by former photodetachment threshold experiments [24]. On deprotonated methyl pheophorbide it is located on C10 (Figure 1a), on pheophorbide and chlorophyll C2 it is located essentialy on the carboxylate of the C7 chain (Figures 1b,c*)*.

We provide here detailed vibrational information on the excited states of pheophorbide and methyl pheophorbide deprotonated anions. We find and analyse intense transitions at low frequencies especially at 368 $cm^{-1}$ close to the proposed coupling vibration for charge transfer, 340 $cm^{-1}$ [4]. This study shows that photodetachment action spectroscopy, associated to simulations, yields detailed information on the excited electronic properties of these deprotonated pigments and by extension on the excited states of neutral chlorophyll pigments.

## Experimental

The photodetachment spectrometer has been detailed in a previous paper [24]. Briefly, gas phase deprotonated anions are produced from a nano ESI ion source (Thermo Fischer, model LCQ DECA XP Plus). They are stored in an octupole ion trap for 100 ms, then pulse extracted and directed into a 3D quadrupole ion trap cryogenically cooled at 9 K. A gas pulse of helium is sent into the trap 5ms before the arrival of the ion bunch. It allows cooling the ions after the gas has equilibrated with the trap. From model developed by Delahaye [25] the ultimate equilibrium temperature should be twice the gas temperature, ~20K. Effective cooling has been checked by the width of rotationally resolved spectra (fluorophenolate) and also by the appearance of clusters of deprotonated pheophorbides with the background nitrogen/oxygen gas entrained by the electrospray process. Such clusters with low binding energies are only formed at low temperatures in seeded beams around 30K. It should be noted that for large molecules rotational broadening becomes a less severe effect compared to vibrational broadening: the rotational envelope of a single vibronic transition *at constant temperature* should decrease with $\approx 1/(B)^{1/2}$, where B is the rotational constant of the molecule. Thereby rotational broadening should be less compared to smaller anions, ~ 20 $cm^{-1}$ [26].

After 80 ms, the cooled ions are pulsed extracted and accelerated to 1800 V in a Wiley-McLaren configuration for time-of-flight mass spectrometry. Then, after pulse extraction from the cryotrap, the ions are spatially (≈0.6 mm) and temporally (30 ns) focused onto the detection devices: the VMI imager for electron detachment and a MCP direct negative ion detector. The anions are photodetached by a 5 ns OPO laser (Horizon Continuum & NT342 EKSPLA) focused on a 0.5 mm diameter, which overlaps the 0.6 mm spatial spread of the 30 ns ion packet travelling at $2.10^4$ m/s. Laser energy is limited to 1-5mJ and neutral density filters can be added. The excitation scheme is sequential where a first photon is resonant with a vibrational state of an electronically excited anion that is photodetached by a second photon of the same color by the same laser pulse.

The images of the velocity distributions of the detached electrons are either transformed by the PBasex procedure [27]and the transformed image is either integrated in a suitable energy range, or integrated without transformation over the whole velocity range of the VMI image..

Spectra are recorded at 0.2 nm (at 500 nm) or 0.5 nm (at 700 nm) intervals for 30 to 40s. This involves a frequency step around 10 $cm^{-1}$ and is smaller than a common rotational envelope of the anions in the range of 20 $cm^{-1}$ at 20K, that is the typical rotational temperature with such a collision cryocooled system[28]. For our systems, which are much larger than most of the spectroscopically observed deprotonated anions, the rotational envelope should be reduced at the same 20K temperature with the inverse of the inertial moment square root of the anion. Thus no increase of the linewitdth is expected for these anions and the narrowest observed transitions contains at least 3 frequency steps. The average laser intensity is recorded in this time interval and used to normalize the electron signal by the square of its value through the spectrum (two photon process). The ion signal is also integrated through the step duration and normalizes the electron signal.

## Calculation methods

All calculations are done using the GAUSSIAN 16 program.[29] The ground state equilibrium geometries as well as harmonic vibrational frequencies have been calculated for the singlet closed shell deprotonated species ($[A\text{-}H]^-$).

We chose to work with the CAM-B3LYP[30] and ωB97X-D3, hybrid exchange-correlation functionals and (/6-311++G**)(d,p) basis sets based on our previous work[31] on deprotonated chlorophyll and pheophytin and the work by Schinke et al[32] on chlorophyllide a and pheophorbide a. We shall only present here the results with ωB97X-D3 which are best adapted to describing the electronic structure of anionic systems, as we used them successfully in former work on the large anion, deprotonated heme[33].

With the results of the DFT and TDDFT mode structure and frequencies in the ground and excited states of the systems under study, we calculated the Franck-Condon factors with the ezFCF program [34,35].

# Results and Discussion

## *Comparison of the neutral and deprotonated pheophorbide pigments*

The electronic spectra of the deprotonated chlorophyll pigments have been reported in a global fashion[36], therefore it was important to examine their electronic absorption spectra in the condensed phase as compared to the neutral protonated molecules. Such spectra were recorded as a function of the basicity in methanol solutions, varied by addition of drops of a base (ammonia or 0.1M NaOH ). This is represented in Figure 2 where the two set of bands in the red (Q) and the blue (B) regions may be observed, which correspond to the well known $\pi\pi*$ transitions of tetrapyrroles[37] originating from two combinations of $\pi$ orbitals of their inner cycle ( vide infra and figure S1 for details) . The reddest band at 680 nm, the Q band, does not shift in frequency at high pH, while the B band at 400 nm displays an isosbestic point at 409 nm confirming the transformation of the neutral methyl pheophorbide into its deprotonated form. In the same basicity conditions, the emission spectra from the Q band have been recorded for both neutral and deprotonated forms and are shown below 660 nm in figure S1. They are quasi identical for neutral and deprotonated species, indicating a probable compensation of the solvent shifts in neutral and deprotonated forms.

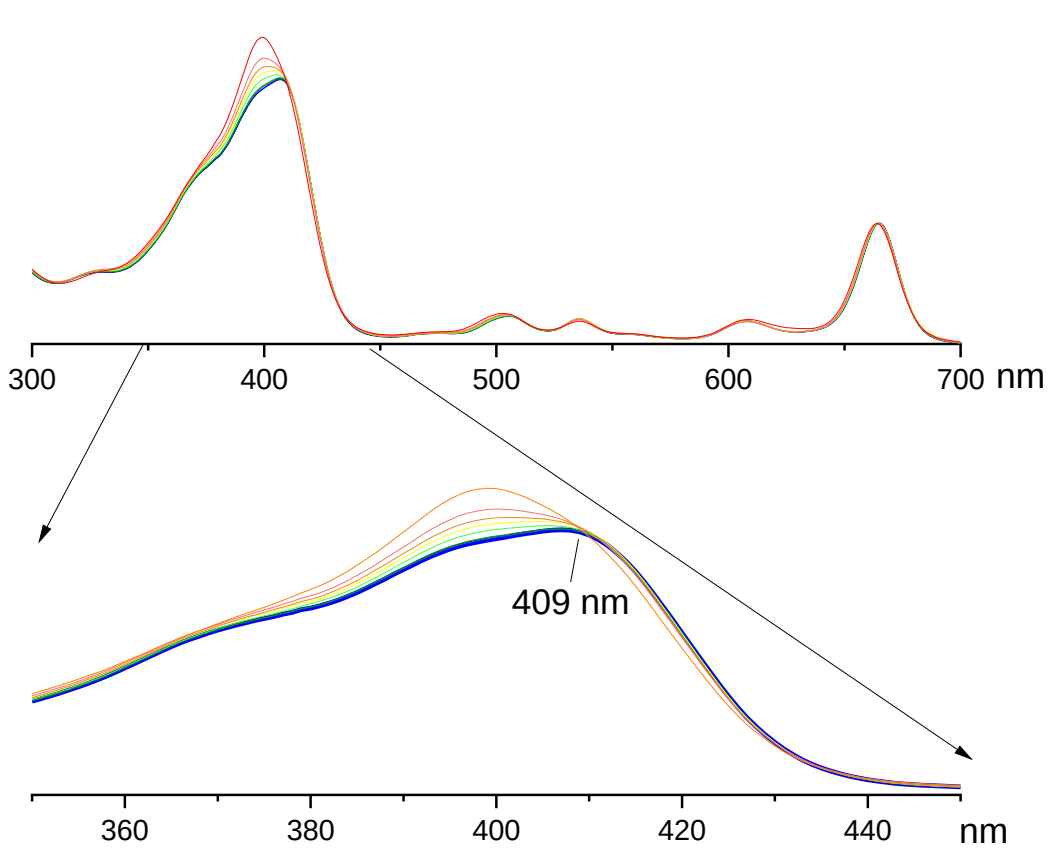

Figure 2: absorption spectra of methyl pheophorbide in methanol at increasing basicity, from neutral (blue trace, lowest) to PH=12 (red trace).

### *Photophysics of the electronic states of the deprotonated anions of pheophorbides*

Action spectroscopy by photodetachment has been applied to cold deprotonated pheophorbide pigments in the gas phase, where they are readily observed and cooled[24]. Action spectra are taken in the 700 - 440 nm region, detecting electrons produced by resonant excitation of the intermediate vibronic states of Q bands as in Figure 3.

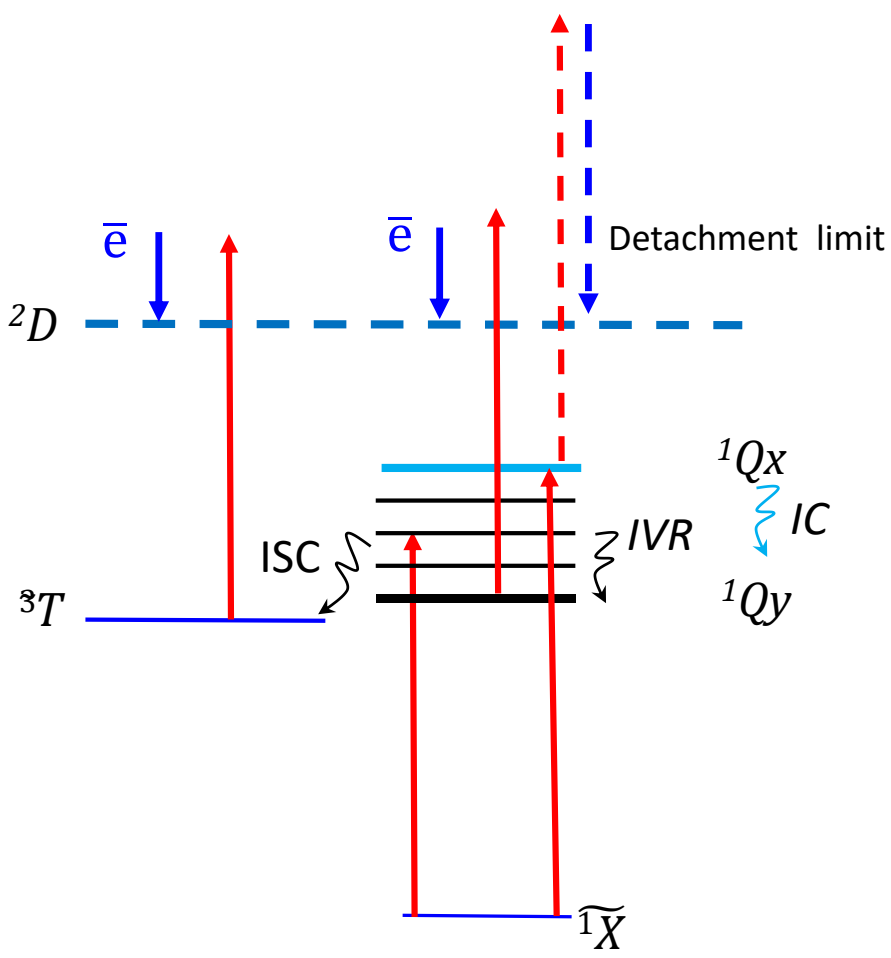

*Figure 3 : Scheme of the resonant two photon electron detachment $S_0$-$S_1$-$D_0$ or $S_0$-$S_1$-$T_1$-$D_0$.*

After resonant excitation of a vibrational state of the Q band of the deprotonated (methyl) pheophorbide, this state can be re-excited directly by a second photon to the detachment continuum *or* follow a relaxation path *before* re-excitation during the nanosecond laser pulse. Relaxation may involve i) intersystem crossing (ISC), ii) vibrational redistribution (IVR) within the Q state or iii) internal conversion (IC).

Intersystem crossing i) occurs on a nanosecond time scale as the fluorescence lifetime occurs on a nanosecond time scale in solutions of the deprotonated ions. Then the triplet anion can be detached, if the photon energy of the second photon allows for. From the neutral triplet (typically 1.3 eV [38]), a photon above 1.4 eV (880 nm) should be required for detachment of methyl pheophorbide with a 2.7 eV detachment threshold and 1.9 eV (650 nm) for pheophorbide with a 3.2 eV threshold. On the other hand, vibrational randomization ii) within the Qy state and internal conversion from Qx iii) redistributes the initial population in far less time than a nanosecond over the other vibrational states of Q yielding a thermalized exited state. This leads to photodetachment from a broad array of vibrational levels in the excited state and to a gradual broadening of the photoelectron spectrum as excitation energy increases (figure S2).

The photoelectron signal $n_e$ is proportional to the $S_0$-$S_1(\nu)$ cross-section $\sigma_1$ to vibrational level v and the $\sigma_2$ is the photodetachment cross-section, : $\frac{dn_e(\nu)}{dt} \propto \sigma_{1(\nu)}\sigma_2 I^2 N_i$ . I is the laser intensity and $N_i$ the number density of anions. The detachment cross-section $\sigma_2$ is strongly influenced by the excess energy of the electron (Wigner correction[39]) and the observed signal $n_e(\nu)$ has been corrected for the $(\text{intensity})^2$ and deprotonated population variations.

In summary, from all states singlet or triplet photodetachment is possible but relaxation decreases its efficiency through a decrease of the available excess energy: it is more favourable for methyl pheophorbide compared to pheophorbide with a 0.5eV lower detachment threshold. *Note that despite relaxation, the first step -Q(ν) characterises resonantly excited level ν.*

### *Observation of the electronic states of the* deprotonated anions of pheophorbides

An ensemble of spectra of deprotonated ions have been obtained with the 5 $cm^{-1}$ resolution of the laser by two photon photodetachment detection. In Figure *4*, the vibrational transitions are resolved with a bandwidth $\geq$ 30 $cm^{-1}$ and rotational broadening is not significant as mentioned in the experimental section.

As can be seen in Figure 4, electronic and vibrational transitions span the whole visible spectrum for methyl pheophorbide and pheophorbide, stopping beyond 450 nm, close to the direct photodetachment threshold of methyl pheophorbide [24]. The spectrum of the pheophorbide deprotonated anion in a basic methanol solution at room temperature has been added in cyan in Figure *4* to show the match between the solution and the gas phase spectra domains: the red bands $Qy_0^0$ and $Qy_0^1$ of the solvated ions being blue shifted in methanol by ~500 $cm^{-1}$ with respect to the gas phase, while Qx bands are red shifted. The two electronic origins Qy and Qx of the deprotonated pheophorbides appear distinctly in the gas phase with extensive transitions. This is quite unexpected for this Qx second excited state, since generally electronic relaxation processes (such as internal conversion), are fast enough to blur vibrational structure.

Qy, $Q_1$ red bands of respectively H- and methyl- pheophorbides display similar shapes and positions. $Q_1$ stands for the red band of methyl pheophorbide and $Q_2$ the green band (vide infra). The relative intensities of the bands and their resolution for the two pheophorbides are different as seen in the b) and c) plots of Figure 4. In the 700 nm region, methyl pheophorbide shows the best contrast compared to pheophorbide. The opposite is true for the Qx (pheophorbide) $Q_2$ (methyl phoephorbide) bands and there is a strong blue shift for pheophorbide with respect to the methylated compound.

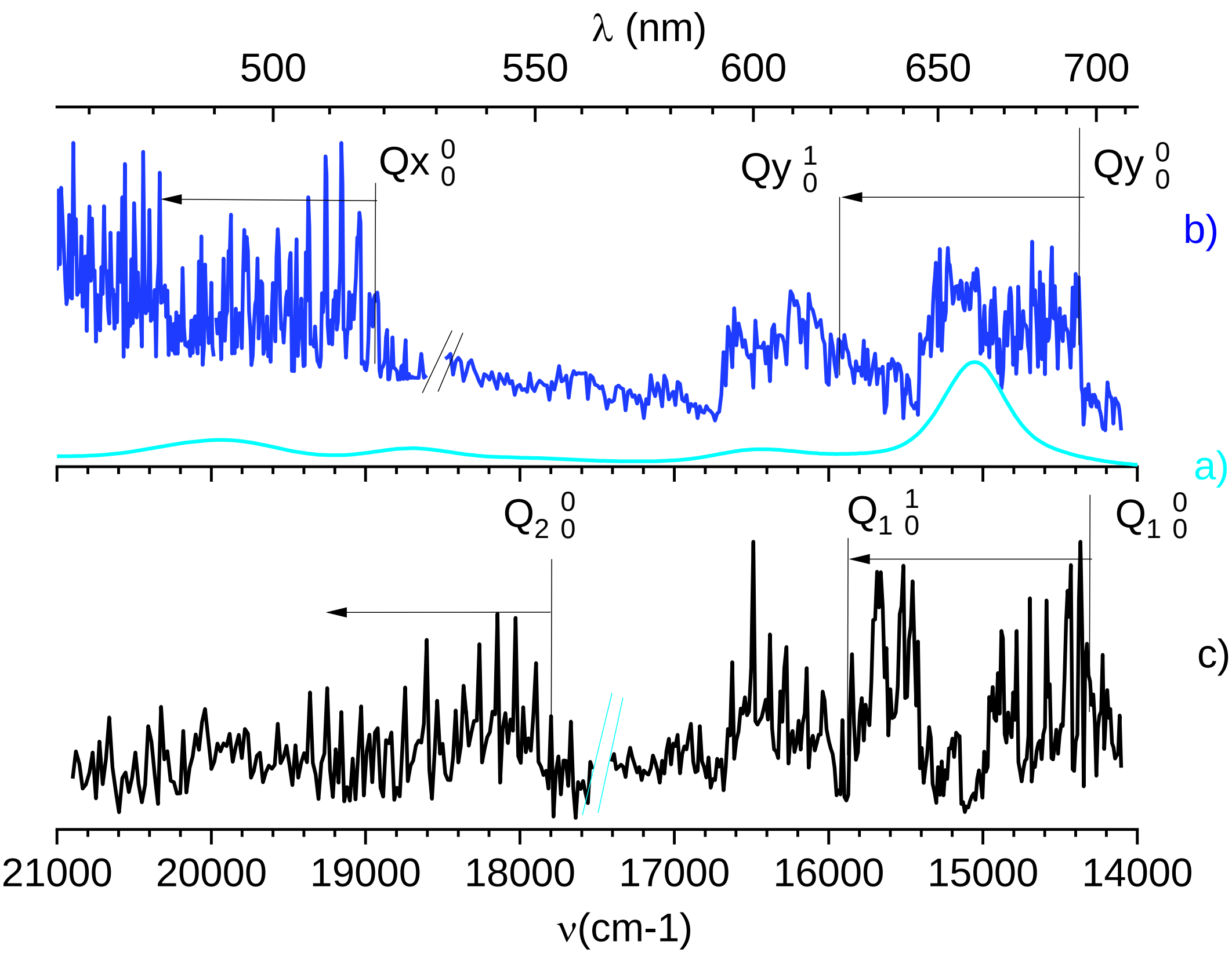

Figure 4 : a) Spectrum of a methanol solution of pheophorbide(cyan), b) 2-photon excitation spectrum of the cooled deprotonated pheophorbide anion with photodetached electron detection (blue), c) 2-photon excitation spectrum of the cooled deprotonated methyl pheophorbide anion (black). // Indicates combination of spectra

### *The electronic origins of the Q bands*

The two electronic origins Qy and Qx calculated for of deprotonated pheophorbide with the DFT hybrid functional ωB97x (/6-311++G**)(d,p) yields a vertical excitation at 661nm for Qy and at 564 nm for its Qx, in reasonable agreement with observations. As we shall further detail, the two states ground and excited have a very similar geometry, hence, the origin Q bands should rise sharply to be the most intense. This appears in the spectra of Figure 4 for pheophorbide and methyl pheophorbide at 695nm.

*Overview of the resolved vibrational structure of the red Qy,$Q_1$* bands

The $Q_y$ / $Q_1$ bands of both deprotonated pheophorbides (H- and methyl) are expanded and compared in Figure 5. We label these red bands for deprotonated pheophorbide as *Qy* and for methyl pheophorbide as $Q_1$, as detailed later. Both red Q bands have almost the same origin and span the same frequency domain 2300 cm$^{-1}$. At both frequency ends of the region, they are very similar, as Figure *5* shows. As the methyl compound is better resolved, we shall focus our attention on its description, since the spectrum of pheophorbide could be contaminated by its deprotomer at C10 [24]. For the methyl compound, a motif of vibrational bands is repeated at +1560 cm$^{-1}$. Curve c) shows a very good match in this region assigning this value as a combination band origin repeating the already observed fundamentals of curve b). Using this comparison, one observes the same spacing for pheophorbide in curve a) and thus the same combination bands.

Globally, one can sort three regions in the spectrum 0-900 , 900-1500, 1500-2300 cm$^{-1}$.

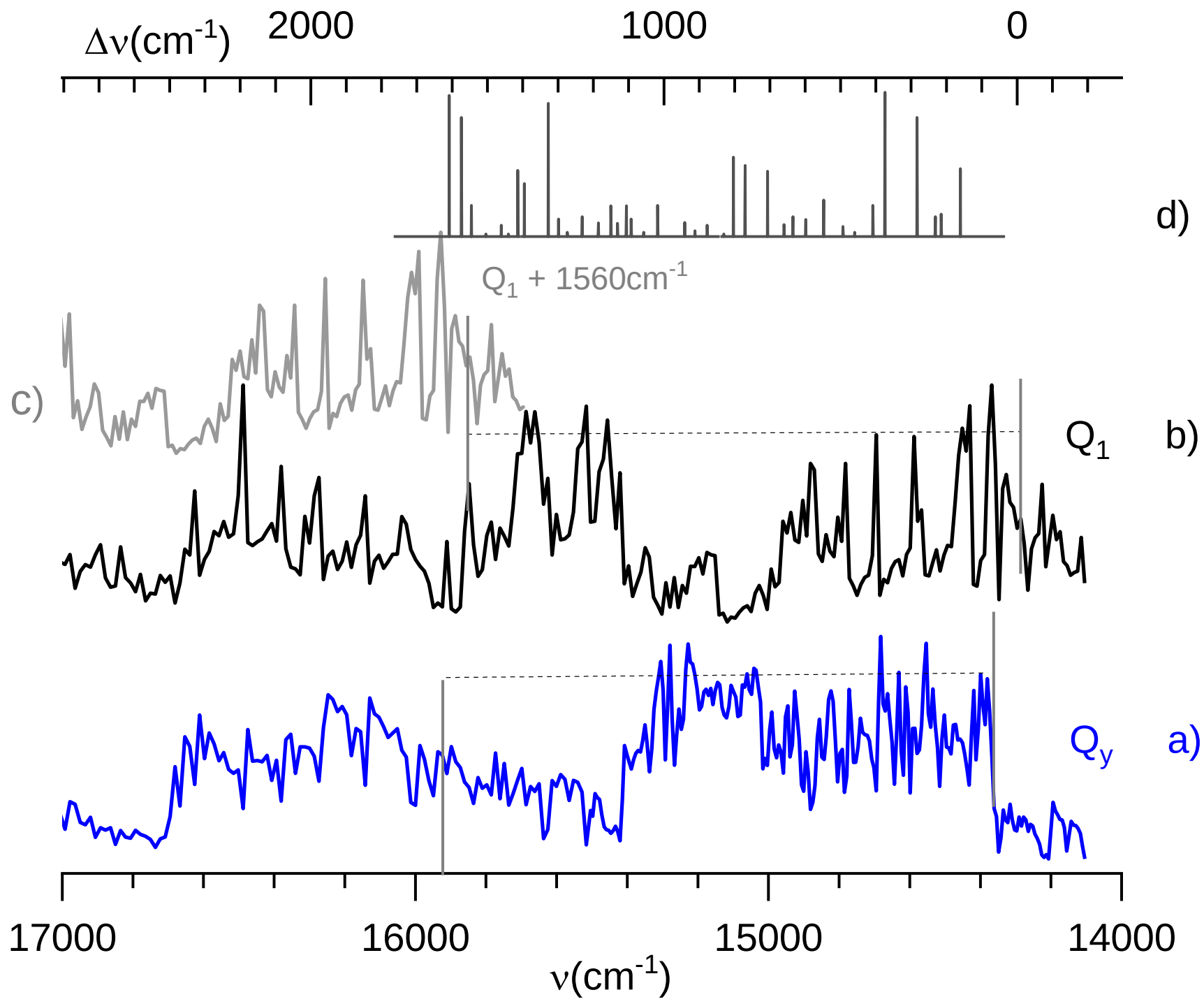

Figure 5 :a) Pheophorbide Qy excitation (blue), b) $Q_1$ excitation spectrum of methyl pheophorbide (black) , c) the same spectrum blue shifted by +1560 cm$^{-1}$ (grey), the global Kékulé group of modes, d) FNLL (frequency line narrowing) excitation peaks at 5K pheophytin (black) extracted from text in reference [40]

The description of these $Q_1$ vibrational bands follows the works of Rebane et al. [13], Avarmaa [6] and Ratsep[40] on cold solid solutions of neutral chlorophyll pigments. We compare our main peak positions with those obtained with the excitation of site selected neutral pheophytin in Table *1* and in Figure *5* d) .

Strikingly within the 0-600 cm$^{-1}$ origin region, we find a good match between our curve b) and the fluorescence line narrowing (FNL) excitation measurements of *pheophytin* by Rebane et al[13] (curve d) . Following his analysis, we find here for methyl pheophorbide two intense transitions at $Q_{1}{}_{0}^{0}$+261 cm$^{-1}$ 'symmetric skeletal' and +369 cm$^{-1}$, as for pheophytin [40] (Table 1). These vibrations also appear in the emission spectrum of pheophytin by Ratsep[40]. The specificity of our measurements is to show the presence of intense very low frequency vibrations (0-200 cm$^{-1}$) listed in *Table 1* : 41, 104, 198 cm$^{-1}$ for deprotonated methyl pheophorbide. One remarks that these latter frequencies differ between deprotonated pheophorbide and methyl pheophorbide. Indeed, these modes contain motion of the C10 chain that is bound differently in the two anions to the macrocycle with the charge on C10 for methyl pheophorbide.

Beyond 1500 cm$^{-1}$, combination bands are built upon a 1560 cm$^{-1}$ fundamental vibration (Figure *5* b) where they are compared with the origin bands in Figure 5c. Also, for pheophorbide in Figure *5*a, the same series of combination bands can be observed.

In region 1000-1500 cm$^{-1}$, although the positions of the deprotonated bands parallel those of neutral pheophytin observed by Rätsep et al.[40], their intensities are quite different. We shall rely for a detailed analysis on the calculations of band positions from harmonic modes calculated by DFT and on intensities from the Franck-Condon calculations with the ezFCF program[34,35].

*Table 1 : Major vibronic transitions above the $Q_{1}{}_{0}^{0}$ origin for neutral pheophytin, neutral pheophorbide (PheoPb), deprotonated Pheophorbide (PheoPb at C7), methyl pheophorbide (Me PheoPb at C10) experimental and calculated. The table contains the frequencies of the transitions that are assumed to pertain to the same vibrational mode as calculated (see text, with a 0.96 global scaling). A brief description of the main movements (with significant Franck- Condon intensity) is extracted from a representation with the Gaussview program.*[29]

| *Neutral Pheophytin Qy state* [40] | *Raman of Neutral PheoPb* | *Deprot. PheoPb* | *Deprot. Me PheoPb* | ν (Qy) *Calc MePheoPb* | *Mode Number Mel PheoPb* | Mode Description |
|---|---|---|---|---|---|---|
| | | 67 | *41* | 43 | 6 | Rock(C7 C10 groups) wag (C3 group) T(C1 C2 groups) |
| *127* | | 139 | *104* | 90 | 15 | global out of plane macrocycle in plane deformation), T (C7 C10 C4 ) rock ( C1 C2 groups) wag (C3 C4 C5 groups) |
| *181* | | 166 | | 153 | 26 | global out of plane macrocycle deformation, T (C7 C10 ) rock ( C1 C2 groups) wag (C3 C4 C5 groups) |
| *198* | | | *198* | 194 | 33 | global macrocycle out of plane deformation, T(C1 C2 C10 groups) rock(C5 C8 group) wag(C3 C4 groups) |
| 250 | | 245 | 261 | 280 | 45 | macrocycle puckering Cα, Cβ Cδ + in plane pyrrole IV III movement wag(C1 C2 C3 C4 C5 groups) T(C10 C7 groups) |
| | | 298 | 304 | | | |
| 341 | | 368 | 369 | 330 | 51 | Cα, Cβ Cδ, 33, pyrrole V compression wag(C10,C5 groups), T(C7 group) |
| 375 | | 388 | | 346 | 52 | Global in plane macrocycle deformation, pyrrole I V compression II deformation in plane twist, T(C10 C7 groups), rock (C1 C2 groups) wag(C4 C5 group), δ(C10 C9 O) |
| 426 | | | | | | |
| 459 | | 455 | 455 | | | |
| 514 | | 520 | 510 | 496 | 62 | Concerted Pyrrole II in plane deformation I III IV V out of plane deformation rock(C1 C2 C3 C4 groups), wag(C5 group) T(C7 C10 groups) |
| 565 | | | 554 | 555 | 66 | In plane concerted rotation of pyrroles II III and pyrrole V deformation rock(C3 C5 groups ) δ(CαH, CβH,NaH,NbH) wag( C8 C1 C2 groups) T( C2 C4 C7 C10 groups) |
| 601 | 603 65 | 586 | | | | |
| 626 | | 636 | 610 | 616 | 70 | Global puckering (NaH NbH C1 C2 C5 C8 groups) |
| 673 | 664 70 | 681 | 666 | | | |
| | | | 700 | 716 | 84 | In plane deformation of pyrroles IV I II, wag( CδH CβH NaH ) Rock (C2 C3 C4 C5), scissoring C8 , T( C2 C7C10) δ(C10 C9 O) |
| 737 | 739 81 | 737 | | | | |
| 770 | | | | | | |
| 797 | 791 89 | 808 | | | | |
| 844 | | 848 | 848 | 841 | 98 | In plane concerted Pyrrole IV III V I II deformation T(C7 C10 C4 groups) rock(C5 group C1a C2b C3a) |
| 879 | 889 97 | 873 | 871 | 881 | 102 | all pyrrole in plane deformation centered on IV I III V II, T( all peripherals) |
| 908 | 916 101 | 930 | 941 | | 105 | pyrrole IV I in plane deformation δ (C12 C α C13) -δ (C14 Nc Cβ) rock (C1 C2b C5 groups) δ (NaH NbH) T(C7, C8 groups) |
| 984 | | 997 | | | | |
| 1024 | | | 1023 | 1007 | 113 | Concerted deformation pyrrole III IV V, wag(C8 group) Rock(C5 group), T(C7 C10 groups) δ(NaH,NbH, CδH) |
| Cd1059 | | | | | | |
| 1073 | | | | | | |
| 1098 | | | 1094 | 1096 | 125 | global in plane pyrrole deformation (IV III II I V) δ(C8H CδH NaH NbH) rock(C1,C2,C3,C5 groups) |
| 1117 | 1164 128 | 1148 | 1129 | 1127 | 130 | global in plane pyrrole deformation (IV III II I V) δ(C8H CδH NaH NbH) |
| 1152 | | | | | | |
| 1198 | | | 1189 | 1211 | 143 | ν(Cδ-C11, C12-Cα) combined with pyrrole I , II deformation , pyrrole III &IV antisymmetric deformation with pyrrole V δ(C8H CαH CβH CδH NaH NbH) |
| 1240 | | | | | | |
| 1265 | | 1292 | 1274 | | | |
| 1294 | | | | | | |
| 1362 | 1360 152 | 1387 | 1335 | 1357 | 161 | Global macrocycle in plane breathing through pyrrole I II III IV V stretch δ (CβH CδH CγH),H(C7)) umbrella C1 C2 C3 C4 C5 C8 |
| 1380 | | | | | | |
| 1407 | | | | | | |
| 1427 | | 1387 | 1434 | 1438 | 181 | pyrrole I, II stretch combined with ν(C12 Cα) scissoring (C1 C2 C4 C5 C8 C7a) δ(CαH ,CβH NaH) |
| *1471* | | | *1459* | 1461 | 190 | Global Kekulé: in phase stretch pyrroles I II IIII with C12 Cα, C14 Cβ C17 Cγ compression , δ (CαH, CβH, NaH, NbH) umbrella C3 C4 C5 groups |
| | | | | | | |
| *1512* | | | *1521* | 1561 | 195 | Global Kekulé in phase pyrrole I III out of phase II IV V, δ (CαH,CβH, CδH NbH NaH ) rock(C5 group) C 9=O group |
| *1540* | | | | | | |
| *1575* | | 1598 | *1584* | 1592 | 197 | Partial Kékulé pyrroles II III V with ν( Cα-C13 , C14 -Cβ–C15) νC9= O T(C3 C4 C5 groups) |
| | | | *1712* | 1666 | 202 | C 9=O Cycle V stretch |
| | | | *1777* | | | |
| | | 1805 | *1816* | | | |
| | | | *1868* | | | |
| | | | *1907* | | | |
| | | 1922 | *1947* | | | |
| | | 2012 | *2054* | | | |
| | | | *2080* | | | |
| | | 2151 | *2162* | | | |
| | | | *2216* | | | |
| | | 2286 | *2299* | | | |
| | | | *2396* | | | |
| | | | *2452* | | | |
| | | | *2509* | | | |
| | | | *2565* | | | |

### *Second electronically excited states: the resolved vibrational structure of the Qx and $Q_2$ bands*

We were surprised to observe well resolved bands in the Qx region of deprotonated pheophorbide, the second allowed electronic excited state of pheophorbide and to a lesser extent in the same region $Q_2$ for methyl pheophorbide (Figure *6*).

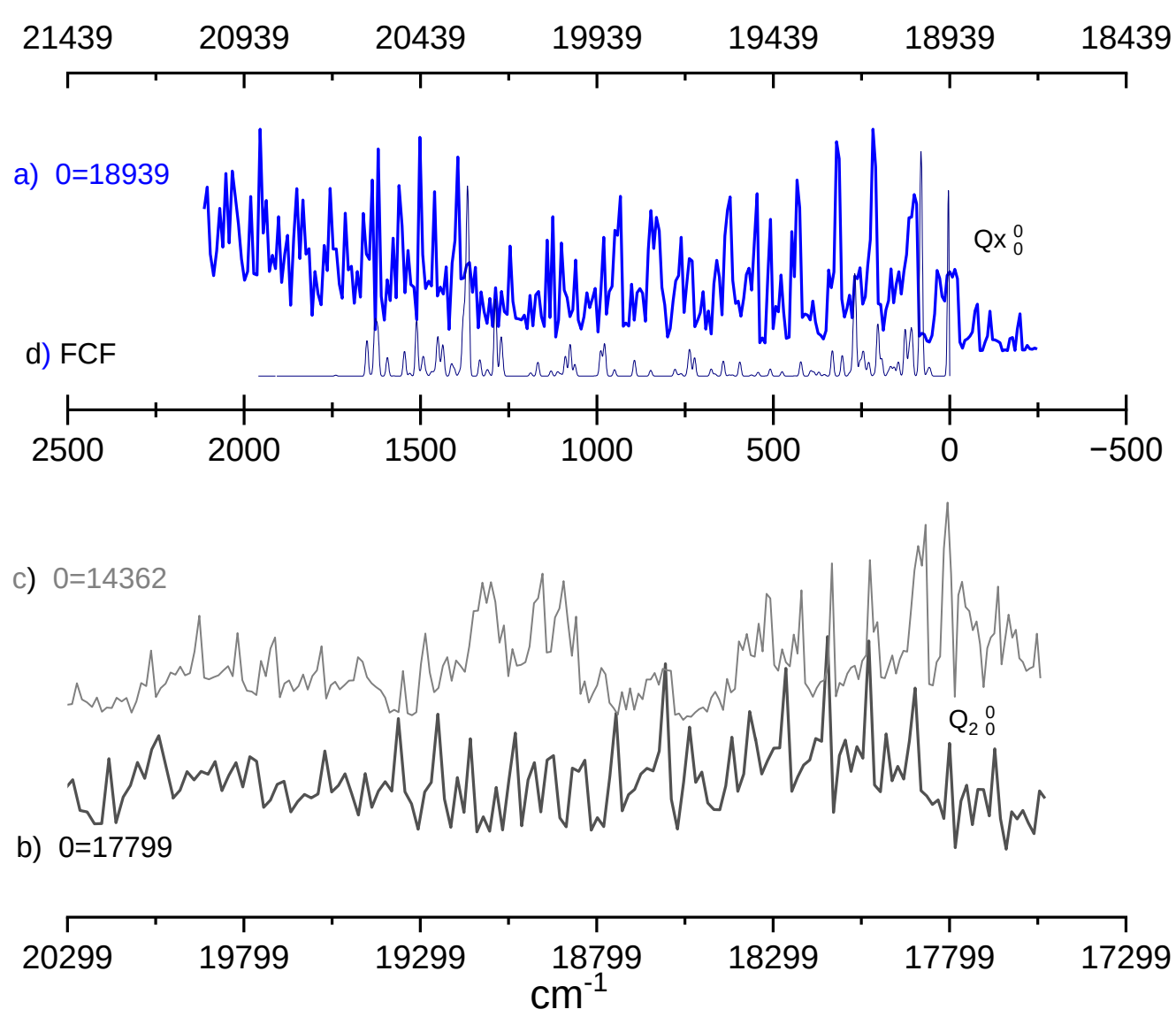

Figure 6 a) Qx spectrum of deprotonated pheophorbide (blue), b) $Q_2$ spectrum of deprotonated methyl pheophorbide shifted by 1140 cm$^{-1}$ with respect to Qx of deprotonated pheophorbide for comparison (black). c) $Q_1$ *origin* spectrum of methyl pheophorbide shifted by 4577 cm$^{-1}$ to 18939 cm$^{-1}$ for comparison(grey), d) Franck Condon simulation of the second excited state of deprotonated pheophorbide (see text),

Strikingly, the bands of the second excited state Qx, ($Q_2$) of deprotonated pheophorbide (methyl) are narrow (<20 cm$^{-1}$) and well resolved, which is exceptional for this second excited state in a molecule of this size. No vibrationally resolved spectrum has been obtained with line-narrowing methods[40] beyond 1600 cm$^{-1}$ in chlorophyll pigments encompassing the Qx region. Besides, with increasing vibrational excitation, broadening does not show up, only the complexity of overtones and combination bands appear close to the detachment threshold. When comparing Figure *6* b and c, no apparent increase of the linewidth is observed for this second excited state as compared to that of the first for deprotonated pheophorbide. Thus the linewidth broadening due to Qx / Qy relaxation is much smaller than the observed linewidth 20cm$^{-1}$. Nevertheless, relaxation occurs from the state reached by Qx excitation: photoelectron spectra of deprotonated pheophorbide were recorded with $Qx^0_0$ excitation and showed no electron energies that would indicate a direct (resonant) detachment from Qx, rather a broad distribution extending to 1eV, well below the energetic limit 1.5eV, as schemed in Figure 3. Thus, prior to photodetachment the molecule has undergone internal conversion to Qy faster than the laser duration - 5ns. .

These Qx bands of deprotonated pheophorbide and $Q_2$ for methyl pheophorbide are compared in curves a) and b) of Figure 6. Curve b) has been blue shifted by 1140 cm$^{-1}$ for this comparison. There appears for both anions the same general vibrational structure with two groups of transitions separated by 1425 cm$^{-1}$. The subbands close to the $Qx^0_0$ origin display a very similar structure, which is also reminiscent of the $Q_{1_0}^{0}$ red origin transitions for methyl pheophorbide shifted by 4577 cm$^{-1}$, as represented in curve c). In the same way, as shown for $Q_{1_0}^{0}$ methyl pheophorbide, the origin pattern of Qx pheophorbide is repeated at +1425 cm$^{-1}$ showing a combination band. The main peaks are collected in Table 2. The observation of these bands might serve as a reference for their identification in the metallated deprotonated pheophorbide compounds where they interact with the lowest Qy transitions due to their closer energy.

*Table 2* Major vibronic transitions above the $Q_{x_0}^{0}$ origin for deprotonated pheophorbide and $Q_{2_0}^{0}$ origin for deprotonated methyl pheophorbide. The most intense FCF values for pheophorbide are included . We also compare in the supplementary section, table S2 , the correspondence of the modes for Qx and Qy in deprotonated pheophorbide.

| Deprotonated Pheophorbide $Q_x$ bands | FCF Pheophorbide Qx bands a) | Deprotonated Me Pheophorbide $Q_2$ bands |
|---|---|---|
| | 79 | |
| 117 | 108 | 167 |
| | 125 | |
| 220 | 205 | 235 |
| | 249 | 286 |
| | 273 | |
| 332 | 339 | 337 |
| 407 | | 371 |
| 445 | | 475 |
| 529 | | 528 |
| 560 | | |
| 637 | | 616 |
| 660 | | |
| 730 | | |
| 746 | 756 | 776 |
| 832 | | |
| 880 | | |
| 960 | | |
| 1000 | 1004 | |
| 1024 | 1015 | |
| 1145 | 1105 | |
| 1260 | | |
| 1293 | 1306 | |
| 1359 | | |
| 1401 | 1404 | |
| 1435 | | |
| **1460** | 1477 | |
| | 1492 | |
| **1519** | | |
| **1536** | | |
| **1561** | 1553 | |
| **1612** | 1588 | |
| **1655** | 1674 | |
| | 1698 | |
| **1732** | | |
| **1750** | | |
| **1775** | | |
| **1802** | | |
| **1854** | | |
| **1880** | | |
| **1933** | | |
| | | |

*a) (peak values=10%max)*

### Comparison of deprotonated anions with the neutral protonated pheophorbides

The fluorescence (Figure S1) and absorption (Figure *2*) measurements of pheophorbide and methyl pheophorbide in methanol / water solutions show that deprotonation modifies little the electronic spectral properties between neutral and deprotonated species, though increasing pH allows the full deprotonation of the neutral. The inspection of the wavefunctions (Kohn-Sham) of deprotonated and protonated (neutral) forms shows also the close similarity between neutrals and both anions, see Figure 7.

The electronic transitions can be pictured assuming a local approximate $d_{4h}$ symmetry, with the Gouterman 4-orbital model[37] where ground state orbitals are 2 combinations of HOMO - HOMO-1 (respectively $a_{1u}$ $a_{2u}$ in $d_{4h}$ symmetry). The lowest transitions originate from there to 2 LUMO ($e_{2g}$ in $d_{4h}$ symmetry). Typically, the two low lying allowed transitions Qy and Qx are shown in Figure 7 using the 4 representations of the contributing Kohn Sham orbitals. The latter represent quasi exclusively the contributing orbitals and resemble the ones of Gouterman[37]. In the case of deprotonated pheophorbide at the carboxylate (C7 group), the 4 orbitals are most similar in shape (see figures S3 a and b) and have the same order in the deprotonated and neutral forms. Transitions will have the same order in the anion and neutral.

In turn, for neutral methyl pheophorbide orbitals 160 ($a_{1u}$ type) and 161($a_{2u}$ type) for the HOMO become respectively 161($a_{1u}$ type) and 160 ($a_{2u}$ type) in the deprotonated forms, see Figure 7. This should also exchange the order of the Qy and Qx transitions for deprotonated methyl pheophorbide with respect to the neutral since the transition moment is linked as drawn, to the main orbital contribution. We should therefore label the transitions in deprotonated methyl pheophorbide as $Q_1x$ and $Q_2y$, the index $_{1,2}$ refers to the spectral position and y,x to the polarisation of the transition. When pheophorbide is deprotonated at C10 on cycle V(fig S3c), the same orbital inversion is obtained as in methyl pheophorbide seen in Figure 7 (left) and the same transition moment polarisation is obtained for the relevant transitions.

The transitions calculated with the wB97xD/6-311++G(d,p) functional and basis sets for pheophorbide locates Qy at 1.93 (table S1-b) and 1.88 eV (table S1-a) respectively for the neutral and deprotonated forms (at the carboxylate of the C7 group). For the neutral pheophorbide, the transitions, the shape of the orbitals and the expansion coefficients of transitions Qx and Qy into orbitals 160 and 161 are very similar to those in the calculations of Qu et al. [32] The neutral Qx energy 1.88 eV calculated value for deprotonated pheophorbide matches the present experimental results at 1.79 eV.

For methyl pheophorbide neutral, calculations yield Qy at 1.93eV (table S2-b), while a much lower value 0.9eV is found for the deprotonated form in C10 (table S2-a). The same pattern is observed for pheophorbide when deprotonated at C10 (table S1-c). From the observation of fluorescence at 690 nm in solutions of deprotonated methyl pheophorbide and from our present measurements that show identical $Q_1$ (methyl pheophorbide), Qy (pheophorbide) origins we can assess the position of the lowest singlet state and the contributing orbitals.

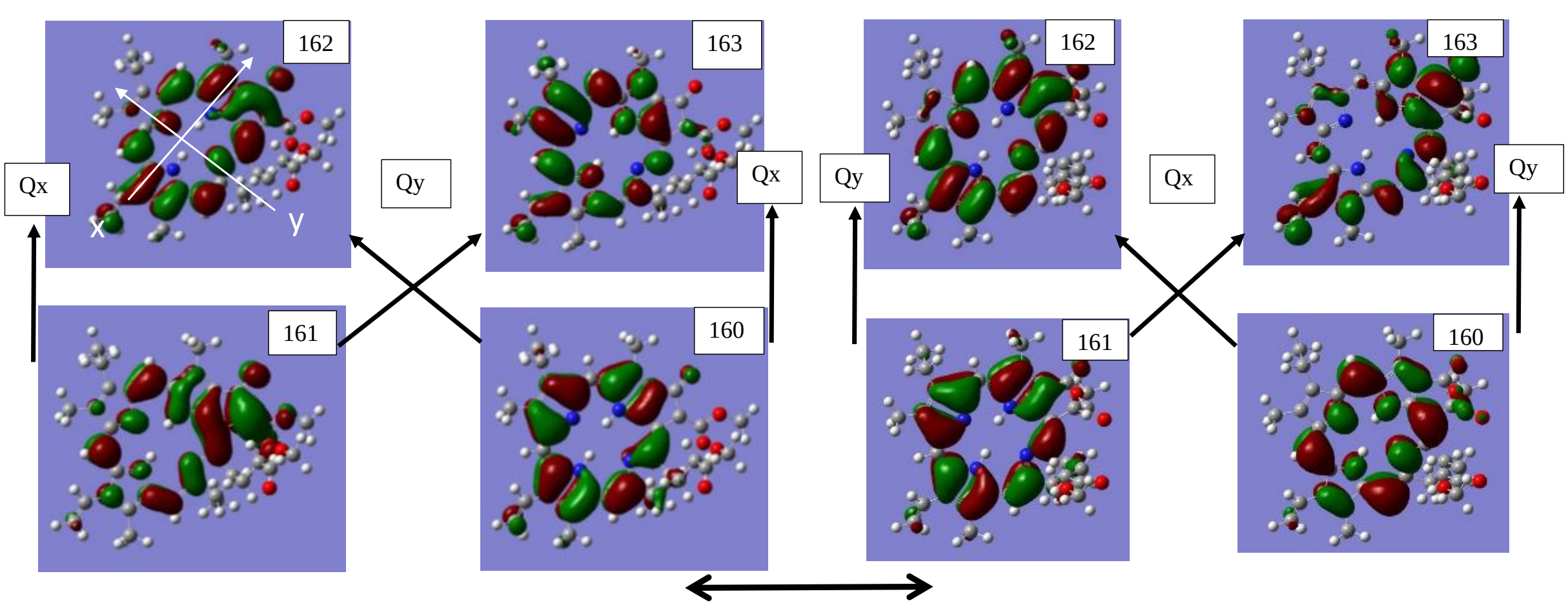

*Figure 7 :Methyl pheophorbide. dominant contributions to neutral orbitals (right) increasing order HOMO-1 HOMO, LUMO, LUMO+1, the same for the deprotonated species (left). On notes the inversion of HOMO's 160 and 161 between neutral and deprotonated species. The coefficients for the functions contributing to the lowest energy transitions are given in table S2 (a and b) and the connections between HOMO's and LUMO's is related to the polarisation of the transitions.*

The main difference between deprotonated pheophorbide and methyl pheophorbide is the presence of three low lying orbitals located on the carboxylate function of pheophorbide (#154,156 157) which do not contribute to the lowest transitions as described by TDDFT (Figure S5). These latter orbitals are also not contributing to the second excited state of deprotonated pheophorbide, which is obtained at 2.29 eV compared to 2.36 eV (experimental).

Thus, the similarity of the electronic distributions in ionic/neutral forms lends to compare the vibrational properties of electronic state of deprotonated species with their neutral correspondents directly found in nature. The comparison of the peak separations in Figure *5* b), d) and *Table 1*. for the excitation of neutral pheophytin by Ratsep et al.[40] shows the match of many transitions with those of deprotonated pheophorbides, with different intensities. Therefore aside from this comparison, electronically excited vibrational transitions of deprotonated anions are more fully described by quantum calculations followed by harmonic analytical Frank-Condon simulations using the ezFCF program[34,35].

Also, an experimental peak may result from the superposition of two modes executing a similar movement at a different location of the macrocycle. For instance, the 4 pyrroles may execute independent or combined movements in different regions of the macrocycle. As an example, listed in table S3 for calculated Qy modes 115 to 119 of pheophorbide relate to 4 different pyrrole stretches within 35 $cm^{-1}$ coupled with different side chains, therefore we have no means or reason to distinguish them. In turn, when the movements become collective, the intensity of the transitions will increase and the assignment can be unique.

We gathered in Table 1 the most intense experimental peaks together with the matching Franck-Condon peaks of the single vibrational excitations (fundamental frequencies) and a descrition of the movements.

### *Resonance Raman spectra of pheophorbide and comparison with deprotonated calculated spectra*

Resonance Raman spectra were recorded in room temperature solutions of pheophorbide and methyl pheophorbide, they are very similar and span a region from 500 to 4000 $cm^{-1}$ . The most intense bands are the concerted pyrrole breathing vibrations around 1500 $cm^{-1}$ and the CH stretches around 3000 $cm^{-1}$. The main region of the Resonance Raman spectrum (0-2000 $cm^{-1}$ ) is represented in Figure 8 and the peaks are listed in table S4 of the supplementary material.

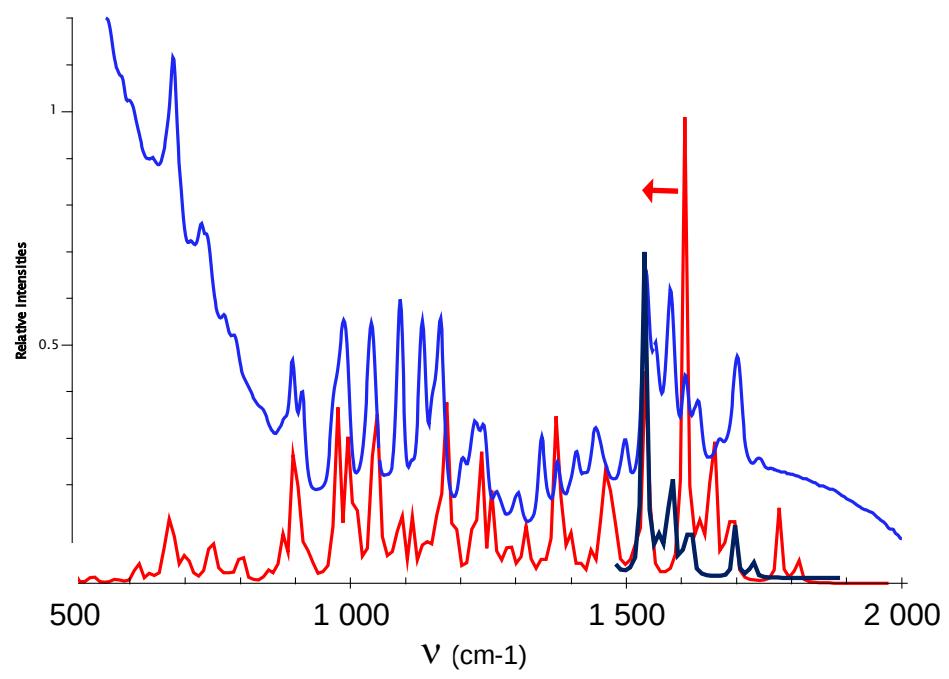

*Figure 8 : Resonance Raman spectrum of pheophorbide blue line. DFT simulation (wB97xD/6-311++G(d,p) scaling factor 0.98) simulation of the Raman spectrum of depronated pheophorbide, as used in the Franck Condon simulations of the electronic spectra, red line. In black, same simulation with a .96 scaling for the Kekulé Bands.*

The Raman spectrum has been calculated for *deprotonated* pheophorbide with the geometry used for the ground state of this molecule in the Franck-Condon calculations. It is reproduced in red in Figure 8 with a 0.98 scaling factor accounting for the anharmonicity. Within the accuracy of these harmonic normal modes calculations, the agreement is excellent between the neutral and the deprotonated systems. This is a further confirmation of the validity of the investigation of the deprotonated species as models as well as their modelling.

### *Description of the $Q_1x$ transitions of deprotonated methyl pheophorbide with FCF simulations*

We have first considered the single vibrational excitations (fundamentals) represented in Figure 9 in blue, excluding the 10 lowest modes ($\nu$<70 $cm^{-1}$). These latter modes involve torsions of the lateral chains at C7 and C10 locations of methyl pheophorbide (see Figure *1*) and the out of plane warp of the macrocycle plane. Above 70 $cm^{-1}$, the movements become essentially in-plane deformations. The FCF intensity of fundamentals drops short around 1600 $cm^{-1}$ above the electronic origin, showing that the most intense fundamentals are located below this value. The assignments of these fundamentals have been consigned in *Table 1*. There one can classify the types of intense vibrations.

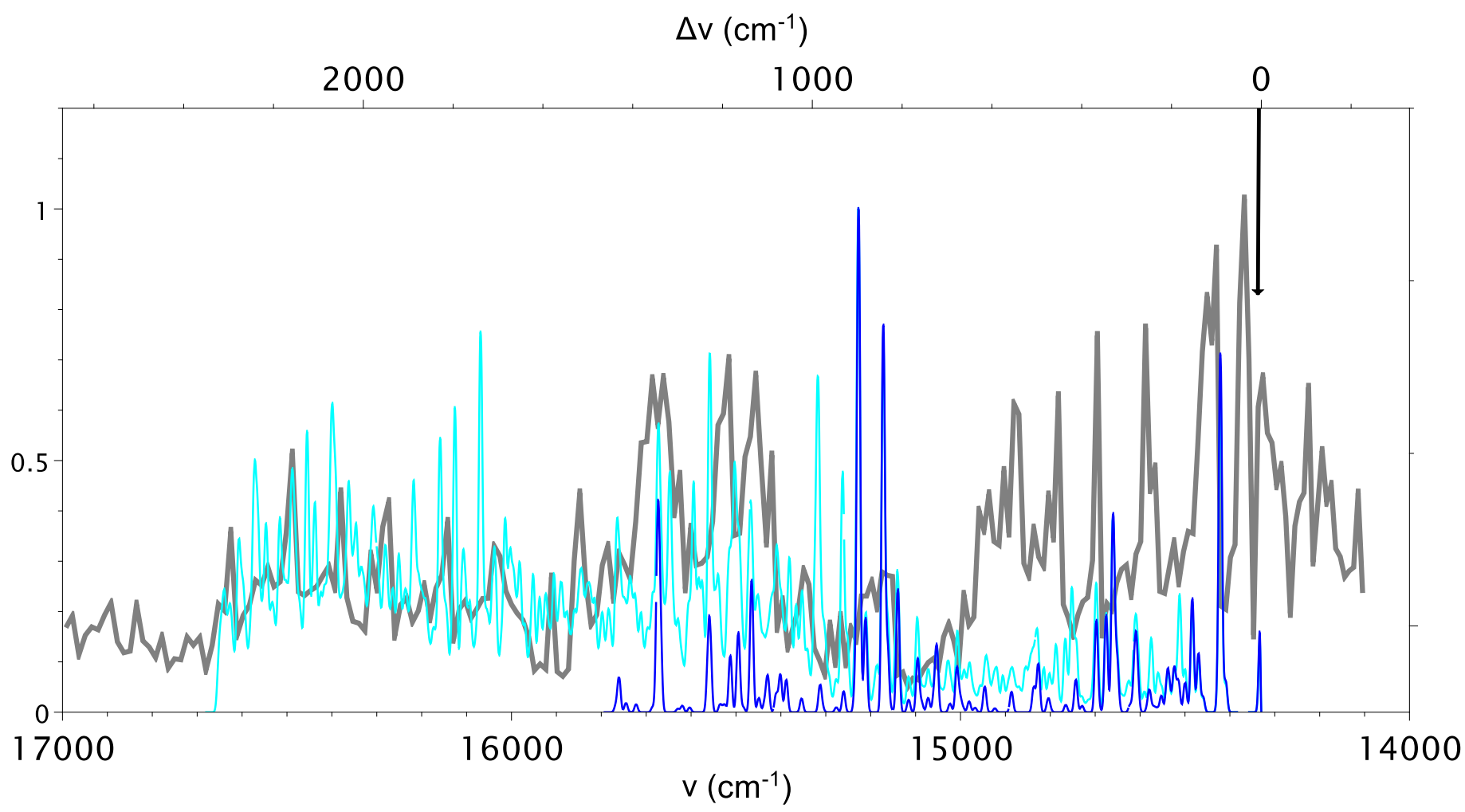

*Figure 9 ;deprotonated methyl pheorbide 2 photon detachment spectrum (black) , FCF simulation (see text) centred on the $Q_1$ 0-0 transition, limited 1) to single excitations of vibrations (blue) 2) excitation of overtones and combination band (up to n= 4) cyan. The scaling factor 0.96. has been used for all vibrations.*

Below 900 $cm^{-1}$ one observes mostly in plane and out of plane global deformations of the molecule. Typically at 280 $cm^{-1}$ a global out of plane puckering of the methyne -CH= bridges (at Cα, Cδ, Cβ,C16 ) combined with in plane movements of the pyrroles that compress the macrocycle matching a 261 $cm^{-1}$ experimental peak in the neutral and anionic forms. At 496 $cm^{-1}$, stiffer deformations set in by strong out of plane deformations of the pyrroles followed by the rocking of peripheral groups (510 $cm^{-1}$ experimental). At 716 $cm^{-1}$ an out of plane global deformation is observed with some experimental activity at 700 $cm^{-1}$. 841 and 881 $cm^{-1}$ modes combine in plane compressions of pyrrole cycles observed at 848 and 871 $cm^{-1}$ respectively.

At 1227 $cm^{-1}$ C=C and C-C stretch combine over all pyrroles and methyne bridges, which probably justifies the intensity of the FCF and the broad experimental peak at 1129 $cm^{-1}$ . There is also in pheophorbide a strong Raman band at 1226 $cm^{-1}$. There appears close to 1357 $cm^{-1}$ and further a combined stretching of the macrocycle with inversion movements of all peripheral $CH_3$'s which matches a broad photodetachment peak at 1335 $cm^{-1}$ and a Raman peak in neutral pheophytin.

At 1461, 1561 and 1592 $cm^{-1}$ a combined alternant compression elongation of double and single bonds (*Kékulé vibrational resonance*) covers half of the macrocycle. These modes can match the 1459/1521/1584 $cm^{-1}$ peaks at the onset of combination bands. Finally, Kékulé modes combine with the C9 =O stretch at 1666 $cm^{-1}$ (1712 $cm^{-1}$ experimental).

Beyond 1560 $cm^{-1}$, Figure *5* c shows unambiguously that a group of overtones built on this Kékulé mode (C=C and C-C alternate stretch) sets in. It is well modelled in Figure 9b, with 4 overtones included in the simulation.  In this FC simulation, the global pattern of the experimental spectrum is well reproduced. One notes that the movements describing the modes in the simulation not only include ‘skeletal modes’ moving the macrocycle but also entrain the peripheral modes making these movements more complex. Indeed, we modelled the absorption of a bare H substituted macrocycle where all groups were replaced by hydrogens (figure S6). It is found simpler yielding groups of bands at 200, 400, 700 and 1500 $cm^{-1}$ (figure S6), however distinct from modelled methyl pheophorbide. As one can see in *Table 1*, most intense movements corresponding to intense bands involve peripheral groups with shifts of frequencies and doubling of modes.

The peculiarity of the present analysis is that it can be transposed to the $Q_2y$/Qx bands that display very similar vibronic patterns, as exposed in table S3

### *Qx bands of deprotonated pheophorbides*

Qx is the second excited state of pheophorbides and spectroscopic information is absent generally for these second electronically excited states for large molecular systems due to fast relaxation processes blurring the spectra.

The Qx bands have been modelled for pheophorbide in a similar way as in the preceding and again, the correspondence between measurements and prominent FCF’s is good as seen in Table 2 and Figure *6* (curves a and d). The examination of the calculated modes shows a parallel between the set of Qx modes and Qy modes for pheophorbide with the aforementioned 3 domain region:  global deformation/ pyrrole cycle movements / combined C=C and C-C stretches culminating with the alternate stretch compression Kékulé motion. We find a quasi exact correspondence of the mode vectors over 90% of the calculated Qx and Qy modes and frequencies values within 10 $cm^{-1}$ up to 1700 $cm^{-1}$, (see correspondence table S3).

In the same manner, the experimental spectra of the $Q_1x$ and $Q_2y$ domains have been compared for deprotonated methyl  pheophorbide and they superimpose over a large region in Figure *6* b), c).

### *Low frequency vibrations in electronically excited deprotonated pheophorbides*

The lowest frequency vibrations of the Qx/Qy states involve global in /out of plane deformation movements of the macrocycle and the torsion of the C7/C10 chains. We identified in the optimized structure of methyl  pheophorbide hydrogen bonds between the C7b H and C10 a O groups. This interaction certainly influences the equilibrium geometries of these chains making in both ground and excited states the Franck Condon simulation difficult in this low frequency region. Therefore, we deem the experimental values for the low frequency vibrations up to 500 $cm^{-1}$ as a very valuable information that is difficult to obtain by other spectroscopies. In addition, the C7 and C10 chains bear partial charges on their oxygen atoms located at the tail of the C7 chain. Thus, the large amplitude movement of this group should exert a large dipole change and a force drawing external charges in electron transfer processes within natural photosystems.

# Summary and Conclusions

We have introduced here a new approach to the spectroscopy of chlorophyll pigments, through their deprotonated anions. This is the photodetachment action spectroscopy of these anions which are isoelectronic to the neutral pigments. We have shown that these systems are valuable models for the study of neutral of pheophytin involved in the primary electron transfer within photosynthetic systems. Here, the spectroscopy of cold deprotonated anions has allowed a vibrational analysis of a wide region of the visible absorption of these pigments in the gas phase in absence of perturbations by the medium. A combined experimental observation and quantum chemical calculations (DFT & FCF) yields a description of the Qx and Qy transitions in these systems comparable with that of natural neutral systems. They have the same electronic distribution, the same spectral extension and very similar vibrational modes for the neutral and deprotonate systems.

Three classes of intense vibrations through optical excitation to the Q bands have been identified:

●global structural deformations at frequencies below 500$cm^{-1}$ : an intense vibration has been identified at 369 $cm^{-1}$ in methyl  pheophorbide that can be related to global puckering of the methyne bridges and may be linked to the 340 $cm^{-1}$ motion driving charge transfer in the reaction centre of PSII[1] .

●Beyond 1300 $cm^{-1}$ there appears a combined C=C stretching and Kékulé type motion at 1500 $cm^{-1}$

● In the intermediate region, local motion of individual pyrroles, in plane and out of plane.

This provides a direct insight in the properties of the natural systems in their electronically excited states.

We found very interesting properties for the second excited state of deprotonated pheophorbide that displays analysable structure uncommon for such a large molecular system. This highly unexpected property might have important consequences on the properties of the Qx bands of other chlorophyll pigments, in that a longer lifetime might allow a greater amount of energy to be transferred resonantly, without dissipation, from these bands.

We have thus far studied demetallated systems and this approach will be applied to the spectroscopy of cold metallated pheophorbides through their deprotonated anions and we propose that deprotonated pheophorbide (metallated and demetallated) can provide a very fruitfull insight in the spectroscopic properties of chlorophyll pigments.

*Acknowledgements:*
This work was supported by the ANR-21-CE50-0028-01 Electrophylle. The computations were performed at the HPC resources from the "Mésocentre" computing center of Centrale-Supelec and Ecole Normale Supérieure Paris-Saclay supported by CNRS and Région Ile-de-France (https://mesocentre.centralesupelec.fr/). BS thanks Samer Gozem (Georgia State University) for advices on the ezFCFprogram . BS and NS thank Aurélien de la Lande from Institut de Chimie Physique (ICP) for fruitfull discussions on DFT calculations and M. Broquier (ISMO) for invaluable scientific advises and discussions.

# References

(1) Romero, E.; Novoderezhkin, V. I.; van Grondelle, R.,Quantum design of photosynthesis for bio-inspired solar-energy conversion, *Nature* **2017**, *543*, 355.
(2) Pavlou, A.; Mokvist, F.; Styring, S.; Mamedov, F.,Far-red photosynthesis: Two charge separation pathways exist in plant Photosystem II reaction center, *Biochimica et Biophysica Acta (BBA) - Bioenergetics* **2023**, *1864*, 148994.
(3) Nguyen, H. H.; Song, Y.; Maret, E. L.; Silori, Y.; Willow, R.; Yocum, C. F.; Ogilvie, J. P.,Charge separation in the photosystem II reaction center resolved by multispectral two-dimensional electronic spectroscopy, *Science Advances* **2023**, *9*, 7190.
(4) Scholes, G. D.; Fleming, G. R.; Olaya-Castro, A.; van Grondelle, R.,Lessons from nature about solar light harvesting, *Nature Chemistry* **2011**, *3*, 763.
(5) Reimers, J. R.; Cai, Z.-L.; Kobayashi, R.; Rätsep, M.; Freiberg, A.; Krausz, E.,Assignment of the Q-Bands of the Chlorophylls: Coherence Loss via Qx − Qy Mixing, *Scientific Reports* **2013**, *3*, 2761.
(6) Avarmaa, R. A.; Rebane, K. K.,High-resolution optical spectra of chlorophyll molecules, *Spectrochimica Acta Part A: Molecular Spectroscopy* **1985**, *41*, 1365.
(7) Rätsep, M.; Cai, Z.-L.; Reimers, J. R.; Freiberg, A.,Demonstration and interpretation of significant asymmetry in the low-resolution and high-resolution Qy fluorescence and absorption spectra of bacteriochlorophyll a, *J. Chem. Phys.* **2011**, *134*, 024506.

(8) Reimers, J. R.; Rätsep, M.; Linnanto, J. M.; Freiberg, A.,Chlorophyll spectroscopy: conceptual basis, modern high-resolution approaches, and current challenges, *Proceedings of the Estonian Academy of Sciences* **2022**, *71*, 127.

(9) Lutz, M.,Resonance Raman spectra of chlorophyll in solution, *J. Raman Spectrosc.* **1974**, *2*, 497.

(10) Mattioli, T. A.; Hoffmann, A.; Sockalingum, D. G.; Schrader, B.; Robert, B.; Lutz, M.,Application of near-IR Fourier transform resonance Raman spectroscopy to the study of photosynthetic proteins, *Spectrochimica Acta Part A: Molecular Spectroscopy* **1993**, *49*, 785.

(11) Gillie, J. K.; Small, G. J.; Golbeck, J. H.,Nonphotochemical hole burning of the native antenna complex of photosystem I (PSI-200), *J. Phys. Chem.* **1989**, *93*, 1620.

(12) Pieper, J.; Voigt, J.; Small, G. J.,Chlorophyll a Franck−Condon Factors and Excitation Energy Transfer, *J.Phys.Chem. B* **1999**, *103*, 2319.

(13) Rebane, K. K.; Avarmaa, R. A.,Sharp line vibronic spectra of chlorophyll and its derivatives in solid-solutions, *Chem. Phys.* **1982**, *68*, 191.

(14) Hughes, J. L.; Conlon, B.; Wydrzynski, T.; Krausz, E.,The assignment of Qy(1,0) vibrational structure and Qx for chlorophyll a, *Physics Procedia* **2010**, *3*, 1591.

(15) Shafizadeh, N.; Ha-Thi, M. H.; Soep, B.; Gaveau, M. A.; Piuzzi, F.; Pothier, C.,Spectral characterization in a supersonic beam of neutral chlorophyll a evaporated from spinach leaves, *J. Chem. Phys.* **2011**, *135*, 114303.

(16) Gruber, E.; Kjaer, C.; Nielsen, S. B.; Andersen, L. H.,Intrinsic Photophysics of Light-harvesting Charge-tagged Chlorophyll a and b Pigments, *Chem.-Eur. J.* **2019**, *25*, 9153.

(17) Gruber, E.; Teiwes, R.; Kjær, C.; Brøndsted Nielsen, S.; Andersen, L. H.,Tuning fast excited-state decay by ligand attachment in isolated chlorophyll a, *Phys. Chem. Chem. Phys.* **2022**, *24*, 149.

(18) Kjær, C.; Gruber, E.; Nielsen, S. B.; Andersen, L. H.,Color tuning of chlorophyll a and b pigments revealed from gas-phase spectroscopy, *Phys. Chem. Chem. Phys.* **2020**, *22*, 20331.

(19) Moca, R.; Meech, S. R.; Heisler, I. A.,Two-Dimensional Electronic Spectroscopy of Chlorophyll a: Solvent Dependent Spectral Evolution, *J.Phys.Chem. B* **2015**, *119*, 8623.

(20) Hughes, J. L.; Conlon, B.; Wydrzynski, T.; Krausz, E.,The assignment of Qy(1,0) vibrational structure and Qx for chlorophyll a, *Physics Procedia* **2009**, *3*, 1591.

(21) Tsujimura, M.; Sugano, M.; Ishikita, H.; Saito, K.,Mechanism of Absorption Wavelength Shift Depending on the Protonation State of the Acrylate Group in Chlorophyll c, *J.Phys.Chem. B* **2023**, *127*, 505.

(22) Yamano, N.; Mizoguchi, T.; Fujii, R.,The pH-dependent photophysical properties of chlorophyll-c bound to the light-harvesting complex from a diatom, Chaetoceros calcitrans, *Journal of Photochemistry and Photobiology A: Chemistry* **2018**, *358*, 379.

(23) Andreou, C.; Varotsis, C.,Light harvesting and photoprotective states in the marine diatom Fragilariopsis sp.: functional implications of chlorophylls c1/c2 in the fucoxanthin–chlorophyll a/c-binding proteins (FCPs), *RSC Advances* **2025**, *15*, 4322.

(24) Soorkia, S.; Muhieddine, A.; Broquier, M.; Poisson, L.; Soep, B.; Shafizadeh, N.,Photodetachment Thresholds of Deprotonated Chlorophyll Pigments and Structural Characterization of Their Deprotomers, *J. Phys. Chem. A.* **2026**, *130*, 1468.

(25) Delahaye, P.,Analytical model of an ion cloud cooled by collisions in a Paul trap, *The European Physical Journal A* **2019**, *55*, 83.

(26) Liu, H.-T.; Ning, C.-G.; Huang, D.-L.; Dau, P. D.; Wang, L.-S.,Observation of Mode-Specific Vibrational Autodetachment from Dipole-Bound States of Cold Anions, *Angew. Chem. Int. Ed.* **2013**, *52*, 8976.

(27) Garcia, G. A.; Nahon, L.; Powis, I.,Two-dimensional charged particle image inversion using a polar basis function expansion, *Rev. Sci. Instrum.* **2004**, *75*, 4989.

(28) Wang, L.-S.,Perspective: Electrospray photoelectron spectroscopy: From multiply-charged anions to ultracold anions, *J. Chem. Phys.* **2015**, *143*, 040901.

(29) Frisch, M. J.; Trucks, G. W.; Schlegel, H. B.; Scuseria, G. E.; Robb, M. A.; Cheeseman, J. R.; Scalmani, G.; Barone, V.; Petersson, G. A.; Nakatsuji, H.; Li, X.; Caricato, M.; Marenich, A. V.; Bloino, J.; Janesko, B. G.; Gomperts, R.; Mennucci, B.; Hratchian, H. P.; Ortiz, J. V.; Izmaylov, A. F.; Sonnenberg, J. L.; Williams; Ding, F.; Lipparini, F.; Egidi, F.; Goings, J.; Peng, B.; Petrone, A.; Henderson, T.; Ranasinghe, D.; Zakrzewski, V. G.; Gao, J.; Rega, N.; Zheng, G.; Liang, W.; Hada, M.; Ehara, M.; Toyota, K.; Fukuda, R.; Hasegawa, J.; Ishida, M.; Nakajima, T.; Honda, Y.; Kitao, O.; Nakai, H.; Vreven, T.; Throssell, K.; Montgomery Jr., J. A.; Peralta, J. E.; Ogliaro, F.; Bearpark, M. J.; Heyd, J. J.; Brothers, E. N.; Kudin, K. N.; Staroverov, V. N.; Keith, T. A.; Kobayashi, R.; Normand, J.; Raghavachari, K.; Rendell, A. P.; Burant, J. C.; Iyengar, S. S.; Tomasi, J.; Cossi, M.; Millam, J. M.; Klene, M.; Adamo, C.; Cammi, R.; Ochterski, J. W.; Martin, R. L.; Morokuma, K.; Farkas, O.; Foresman, J. B.; Fox, D. J. Wallingford, CT, 2016.

(30) Yanai, T.; Tew, D. P.; Handy, N. C.,A new hybrid exchange–correlation functional using the Coulomb-attenuating method (CAM-B3LYP), *Chem. Phys. Lett.* **2004**, *393*, 51.

(31) Diop, M.; El-Hayek, M.; Attard, J.; Muhieddine, A.; Veremeienko, V.; Soorkia, S.; Carbonnière, P.; de la Lande, A.; Soep, B.; Shafizadeh, N.,Chlorophyll and pheophytin protonated and deprotonated ions: Observation and theory, *J. Chem. Phys.* **2023**, *159*.

(32) Qu, Z.-w.; Zhu, H.; May, V.,Time-Dependent Density Functional Theory Study of the Electronic Excitation Spectra of Chlorophyllide a and Pheophorbide a in Solvents, *J.Phys.Chem. B* **2009**, *113*, 4817.

(33) Muhieddine, A.; Soorkia, S.; Attard, J.; de la Lande, A.; Shafizadeh, N.; Soep, B.,Coordination of Deprotonated Ferrous Heme with CO and O2 in the Gas Phase: Influence of Spin-Orbit Splitting and Charge, *J. Phys. Chem. A.* **2025**, *129*, 7087.

(34) Gozem, S.; Krylov, A. I.,"The ezSpectra suite: An easy-to-use toolkit for spectroscopy modeling",, *WIRES CMS, e1546* **2021**.

(35) Wojcik, P.; Gozem, S.; Mozhayskiy, V.; Mukherjee, M.; Baker, C.; Krylov, A. I.,http://iopenshell.usc.edu/downloads.

(36) Saito, M.; Tanabe, T.; Noda, K.; Lintuluoto, M.,Photodissociation of the monoanions of gas-phase chlorophylls a and b in an electrostatic storage ring, *Phys. Rev. A* **2013**, *87*, 033403.

(37) Gouterman, M.,Spectra of porphyrins, *J. Mol. Spectrosc.* **1961**, *6*, 138.

(38) Bhattacharjee, S.; Neese, F.; Pantazis, D. A.,Triplet states in the reaction center of Photosystem II, *Chemical Science* **2023**, *14*, 9503.

(39) Wigner, E. P.,On the Behavior of Cross Sections Near Thresholds, *Phys.Rev.* **1948**, *73*, 1002.

(40) Rätsep, M.; Linnanto, J. M.; Muru, R.; Biczysko, M.; Reimers, J. R.; Freiberg, A.,Absorption-emission symmetry breaking and the different origins of vibrational structures of the 1Qy and 1Qx electronic transitions of pheophytin a, *J. Chem. Phys.* **2019**, *151*, 165102.